\documentclass[11pt,a4paper]{article}
\usepackage{epsfig,multicol}
\usepackage[dvips]{color}
\begin{document}
\begin{titlepage}
\begin{flushright} 
HU-EP-05/35\\
\end{flushright}
\mbox{ }  \hfill hep-th/0508071
\vspace{5ex}
\Large
\begin {center}     
{\bf D-branes in overcritical electric fields
}
\end {center}
\large
\vspace{1ex}
\begin{center}
Harald Dorn, Mario Salizzoni and Alessandro Torrielli
\end{center}
\vspace{1ex}
\begin{center}
Humboldt--Universit\"at zu Berlin, Institut f\"ur Physik\\
Newtonstra\ss e 15, D-12489 Berlin\\[2mm] 
 \vspace{1ex}
\texttt{dorn,sali,torriell@physik.hu-berlin.de}

\end{center}
\vspace{4ex}
\rm
\begin{center}
{\bf Abstract}
\end{center} 
\normalsize 
We collect some arguments for treating a D-brane
with overcritical electric field as a well-posed initial
condition for a D-brane decay. Within the field theoretical
toy model of Minahan and Zwiebach we give an estimate for
the condensates of the related infinite tower of tachyonic
excitations.

\vfill
\end{titlepage} 

\section{Introduction}

Tachyon dynamics in the context of D-brane decay was the subject
of intense studies in recent years \cite{s}.  The main focus has been on
the
field theoretical or string field theoretical aspects of tachyons as
excitations indicating some instability leading to the decay of certain
field configurations. This field theoretical point of view is by far
more adequate than discussions in terms of a particle language including
its conceptual problems with superluminal velocities.

Open strings in the presence of an overcritical 
electric field mainly have been considered
as an ill-defined setting 
(there are however interesting connections with S-branes, 
see for instance \cite{sbranes1,sbranes2}). 
Classically the endpoints of open strings
have to move with superluminal velocity and at the quantum level
there appears a singularity at the critical field strength in the string
analog of the Schwinger pair creation \cite{b, bp}.
On the other hand, looking at String Field Theory in the presence of an 
overcritical electric field presents a picture conceptually not so
different from that with undercritical or zero electric field. 
Instead of one tachyonic
excitation, accompanied by a zero mass and an infinite tower of
stable excitations, one is faced with a stable excitation corresponding to
the former tachyon and the fact that the infinite tower has become tachyonic.
At this level the difference between undercritical and  overcritical electric 
field seems to be only technical.

In the present paper we want to sketch this point of view in some more
detail.
Section 2 is devoted to a discussion of several aspects supporting the
conjecture that the decay of D-branes due to the presence of an overcritical 
electric field can be
a well posed problem in string theory. Of course any attempt to implement
this picture has to handle the serious technical problems caused
by the infinite number of tachyonic excitations driving the D-brane decay.
The level truncation, which turned out to be very effective in the 
undercritical case, is no longer applicable. Some comments
on the treatment of the overcritical 
case within Boundary Conformal Theory are also added. 
In Section 3 we then restrict ourselves to a discussion within a toy model 
introduced by Minahan and Zwiebach in \cite{z, mz}, 
accordingly modified by the presence of the overcritical
electric field.  As a quantitative result we
will get an estimate of the values of all the fields in the infinite
tachyonic tower after condensation.

\section{General considerations}
\subsection{Spectrum}
Propagation of open bosonic strings in the background of a constant
electromagnetic
field has been analyzed in a series of papers \cite{b,bp,ft,n}. The spectrum 
of the masses of a neutral string in the presence of a purely 
electric background has been derived as a particular case in \cite{b,n}, 
where a classical instability has been shown to arise 
above a critical value of the 
electric field. In this case, string modes develop negative squared masses,
and  a tachyonic contribution to the mass comes also
from the motion of the string in transverse directions. 
This classical
instability, which is indeed present both for neutral and for charged strings,
has no analogue in particle mechanics\footnote{For 
charged strings, another kind of instability due to pair-production
dominates at the quantum level
for weak fields \cite{b,bp}. This is the analog of the 
Schwinger phenomenon for particle electrodynamics. 
The derived rate is however zero 
for neutral strings, preventing this kind of mechanism from screening the 
growth in absolute value of the electric field, until it reaches the critical  
value at which the classical instability appears.
On the other side, raising slowly $E$ exposes the 
bosonic D-brane 
to the usual tachyon decay (which is absent in the similar
case of open superstrings). We assume an overcritical electric
field as a given initial condition.}. 
Its appearance is advocated in \cite{b} 
as a signal of the fact that, in a second quantized 
treatment of the theory, the string field would evolve away from the chosen
unstable configuration. 
The spectrum can be reproduced by the formula
\begin{equation}
\label{spectrum}
k_{\mu} G^{\mu \nu} k_{\nu} = - (n - 1)
\end{equation}
where we set  $\alpha' = 1$. The metric to be used is the so called ``open string
metric''
\begin{equation}
\label{openmetric}
G^{\mu\nu} = \left( \frac{1}{g +  B} g \frac{1}{g -  B} \right)^{\mu\nu},
\end{equation}
where $g, B$ are the sigma-model (closed string) 
constant backgrounds \cite{sw}. 
This, in turn, is 
the natural formula arising from the analysis of the conformal dimensions
of the vertex operators of the worldsheet conformal field theory \cite{sw},
and consistently from evaluation of the singularities 
of string scattering amplitudes (see 
\cite{btv} and references therein). 
For our purposes
we will take the closed string metric to be $g = \eta = (- 1, 1, ..., 1)$. 

In the case of a purely electric background, 
without loss of generality one can choose 
the only non-zero components of the antisymmetric tensor to be  
\begin{equation}
\label{Bfield}
B_{0 1} = - B_{1 0} = E .
\end{equation}
The open string metric becomes $G_{\mu \nu} = (- (1 - E^2), 
(1 - E^2), 1, ..., 1)$. One can notice 
the appearance of a critical value for the electric field:
when $E = 1$, the 
open string metric (\ref{openmetric}) 
changes its signature in the $(0,1)$ block. 
Formula (\ref{spectrum}) can be written as
\begin{equation}
\label{spectrumel}
- k_0^2 + k_1^2 + (1 - E^2) k_{\perp}^2 = - (1 - E^2) (n - 1), 
\end{equation}
or, equivalently,
\begin{equation}
\label{spectrumel2}
k_{\mu} \eta^{\mu \nu} k_{\nu} = - (1 - E^2) (n - 1) + E^2 k_{\perp}^2,
\end{equation}
$k_{\perp}$ indicating the momentum component transverse to the electric 
field. 
If one sets $k_{\mu} \eta^{\mu \nu} k_{\nu} = - M^2$, this 
formula 
coincides with what is obtained for example in Sec. 4 of \cite{n}.
Generically, the distance between equidistant levels of the operator $M^2$
is smaller by a factor $(1 - E^2)$ \cite{n}. 
In the overcritical case, then
one has $M^2<0$ (up to the first level due to presence of the zero-point energy
of the oscillators), and a tachyonic contribution coming from the motion of the string 
in the transverse direction
\footnote{We keep in mind that this last contribution to the squared mass is present even for undercritical fields, while only for overcritical fields it gives a tachyonic contribution to the energy, according to the dispersion relation $k_0^2 = k_1^2 + (1 - E^2) k_{\perp}^2 + (1 - E^2) (n - 1)$ derived from (\ref{spectrumel}). The sign of the ``bare''
squared mass term depending on the level $n$ is always reversed in overcritical fields.}.
 
What we can read out of (\ref{spectrumel}), is that what was
a time-like momentum (positive squared mass) in the absence of the 
background, becomes in the
presence of an overcritical electric field a space-like one
(tachyon).
The spectrum of the bosonic open string has therefore been
reversed, and it contains an infinite tower of tachyonic 
modes\footnote{We remark that a possible 
interpretation of this change
of signature from the point of view of the full target space 
could be ascribed to an interchange of the role of space and 
time between the directions $0$ and $1$. However, if one demands now $x_1$
to assume the role of time, thus recovering a standard positive mass 
spectrum for the open 
string tower, then the closed string sector of the theory, whose masses 
are determined \it via \rm
the closed string metric as $\kappa_{\mu} g^{\mu \nu} \kappa_{\nu} 
= - 4 (n - 1)$,
 would have infinite tachyons.}. 

It is also important to notice that a similar situation,
in the presence of an electric background, would occur in superstring theory
\cite{b,btv}, whose spectrum does not originally contain a 
tachyon. 
The appearance of an infinite tower of tachyons has no analog in the 
absence (or in the presence of an 
undercritical) electric field. Therefore, in the following 
we will regard the ensuing classical evolution away from the chosen 
configuration
(which we will suggest to interpret as a decay of 
the electrified  D25-brane) 
as \it due \rm to the overcritical electric field
\cite{b}.

\subsection{D-brane Decay}
The above mentioned instability is accompanied by another effect. When 
interpreting bosonic open string theory as a description of the dynamics of
a space-filling D25-brane, 
the tension of such a brane is naturally
derived from the Dirac-Born-Infeld
action \cite{l}. 
Such a tension, in the presence of a constant electric field background,
is therefore proportional to the Born-Infeld
factor $\sqrt{1 - E^2}$. This
factor becomes imaginary in the overcritical case, which is to say that the
squared mass of the brane becomes negative. 

We can interpret this occurrence in  the open string 
picture. Classically, an electric field stretches open strings
against their internal tension \cite{sst}. 
At the critical value the electric force counterbalances the tendency of
the string to
oscillate, and stretches it to infinity. Beyond the critical value 
the classical tachyonic instability is generated 
\footnote{At the same time the
endpoints of the strings classically acquire a superluminal velocity
\cite{n}.}. This is a signal 
that the nonperturbative string field theory vacuum, whose dynamics is
determined by the open string theory defined on such a background, 
acquires a tachyonic instability. 
The fact that the D25-brane becomes tachyonic has another manifestation in the
superluminal velocity of the T-dual D24-brane \cite{bach}. Equivalently,
this is known to be a (tilted) S-brane, a spacelike object which 
arises in the description of the standard D-brane decay \it via \rm 
the rolling tachyon picture \cite{sbranes1,sbranes2}. In particular,
in \cite{sbranes1} an effective action has been derived for S-branes,
which turns out to be equal to the Dirac-Born-Infeld action times a 
factor $i$.

We suggest to interpret the features of the 
first quantized spectrum discussed in the previous section in the light of 
this physical picture. While it is by now 
clear how one should interpret the presence of the usual bosonic
open string tachyon, which is responsible for driving the D-brane decay
towards the closed string vacuum \cite{s}, here the situation is complicated
by the fact that the brane itself behaves like a tachyonic soliton
background of String Field Theory. 
This is precisely because, while the nature of the bosonic string tachyon 
is related to string zero point 
quantum oscillations without classical analogue, the tachyonic tower 
described in the previous section reflects 
the classical instability of the tachyonic brane. 
What we propose is therefore
that, still, the idea of describing the decay through the rolling of the 
tachyonic excitations living on the worldvolume is applicable, with the 
natural difference being represented by the presence of an infinite
number of tachyonic states. 
The practical way to do this will be described in the next section.
 
\subsection{String Field Theory and BCFT}
The main tool to study the rolling of the tachyon 
has been Cubic 
String Field Theory \cite{w}, 
in connection with methods of Boundary Conformal Field Theory \cite{s}.
On one side, the presence of a minimum of the String Field Theory potential 
corresponding to the absence of perturbative open string excitations 
has been firmly established in the level truncation scheme. On the other side, 
an exactly marginal boundary perturbation to the world-sheet action for
the tachyon profile of the form
\begin{equation}
\label{bcft}
\lambda \int_{\partial \Sigma} \cosh [X_0 (t)] dt
\end{equation}
has been used \cite{s} to describe the time evolution of the system and to
derive the related stress-energy tensor during the decay. In the presence 
of an 
undercritical electric field, the procedure has been generalized in 
\cite{mukho}. The final products of the decay
include in this case  
the additional presence of fundamental string charges in the tachyon
vacuum.

The idea is therefore to make use of the  available formulation of String 
Field Theory in the presence of antisymmetric backgrounds, 
this time evaluated 
for an overcritical electric field. We still expect it to be the instrument  
which describes 
the nonperturbative decay of the original unstable configuration,
through the evolution of the infinite tower of tachyons.
However, 
if one tries to apply the above mentioned  
strategy to this new situation, one soon realizes the scarce
suitability of the level truncation scheme\footnote{See \cite{carlo} for considerations on critical electric fields in Vacuum SFT.}. Its good approximation relies
on the fact that all the tower of positive mass excitations is stable around the origin
(perturbative vacuum), and even if the higher spin fields assume non-zero
vacuum expectation values at the closed string minimum \cite{ks}, it is 
conceivable that these values will not differ too much from the original 
stable point at zero (this appears also
as a feature of the toy model 
used in \cite{z,mz} on which we will elaborate in the next 
section). Now instead, all of them are
tachyonic at the origin, and one will expect all of them to substantially
move away. Neglecting them all but a finite set, or in other 
words setting them to zero starting from
a certain level on, does not look \it a priori \rm like a good approximation.

On the other hand, since all the tower of states is rolling down, one would 
need like an infinite sum of boundary perturbations of the type (\ref{bcft})
for the higher spin fields, with the suitable modifications in order 
to account for the electric field.

The boundary perturbation (\ref{bcft}) was originally 
introduced by inspection of the linearized String Field Theory equations of 
motion. If one would like to proceed in analogous way, 
one has to look at solutions
of the linearized equations of motion for the higher spin fields of the tower,
in the background of an overcritical electric field.
The main problem is the inclusion of all the tower, and we will sketch here
only the proposal for the first state. 
For this purpose, we 
can follow the treatment reported in \cite{lv}
to determine the condition for conformal invariance
of the boundary coupling to the 
spin two field at the linearized level in the background field. 
Adopting their parameterization and choosing the gauge
where the Stueckelberg field is set to zero, one obtains the term
\begin{equation}  
\label{labasbcft}
\int_{\partial \Sigma} A_{\mu \nu} (X) \, \partial^a X^{\mu} \, 
\partial_a X^{\nu} \, dt
\end{equation}
with the following conditions on the symmetric tensor $A_{\mu \nu} (X)$:
\begin{equation}
\label{transverse}
(\partial^{\sigma} \partial_{\sigma} - 1) A_{\mu \nu} = 0 \, \, \, ; \, \, \, 
\partial^{\mu} A_{\mu \nu} 
= 0 \, \, \, ; \, \, A_{\mu}^{\mu} = 0,   
\end{equation}
where all indices are contracted with the open string metric. 
Taking this into account, and making a spatially 
independent ansatz, one finds that
\begin{equation}
\label{solution}
A_{\mu \nu} (X^0) = a_{\mu \nu} \cosh (X^0 \sqrt{E^2 - 1})
\end{equation}
is a solution, provided $a_{\mu \nu}$ is symmetric, purely spatial and 
traceless. 
Inserted in (\ref{labasbcft}), the solution (\ref{solution})
provides a generalization of the boundary deformation (\ref{bcft})
suitable for the new case. The polarization tensor $a_{\mu \nu}$ would have
the same \it status \rm as the parameter $\lambda$, the initial value of the
tachyon. 

This will have to be supplemented with all the
remaining boundary couplings 
for the higher spin states. However, it can already be studied
as a prototype in order to gain insight about the features of the
boundary state resulting from such kind of perturbations, which will in turn 
determine the type of vacuum obtained at the end of the decay.

\section{A Toy Model}

In \cite{z, mz} toy models for the standard 
bosonic string tachyon condensation were studied, whose 
features simulate the behaviour found 
in String Field Theory for the true rolling tachyon. The 
idea is to study the lump solution of an effective field theory  
for the tachyon field. 
Fluctuations around this nonperturbative solution are determined by solving 
a Schr\"odinger equation of a known type, whose spectrum can be exactly 
computed resulting in an infinite
tower of scalar excitations of increasing mass. 

This situation has a close resemblance with the case of
the String Field Theory obtained by
quantization around the vacuum provided by a D-brane. 
In particular, the presence of a 
tachyon signals the instability of such a configuration, and the decay 
towards a stable minimum can be studied by 
looking for the minima of the 
obtained multiscalar potential. The main advantage is that in this simplified 
case one already knows the minimum of the original potential, and one can fix 
the values of the infinite set of scalar fields in the static case by simply 
requiring that summing the fluctuations to the lump profile produces 
the minimum of the original tachyon field. One can therefore treat
simultaneously the {\it whole} tower of states, which in the case
of an overcritical electric field is
a strict requirement. 

The technical procedure, in the case of absence of electric field, 
is briefly summarized in what follows. One considers 
the action for the tachyon field $\phi$ of a D25-brane 
\begin{equation}
\label{cubic}
S = {{1}\over{g_o^2}} \int d^{25}y \, dx \bigg[-{{1}\over{2}} \partial_{\mu} 
\phi \partial^{\mu} 
\phi - {{1}\over{2}} {(\partial_x \phi )}^2 - V(\phi)\bigg],
\end{equation}
where $y^\mu=(t,\vec{y})$ and  $g_o$ is the open string coupling constant. 
The potential has the form shown in Fig.1: 
\begin{figure}
\centerline{\includegraphics[width=4.8cm]{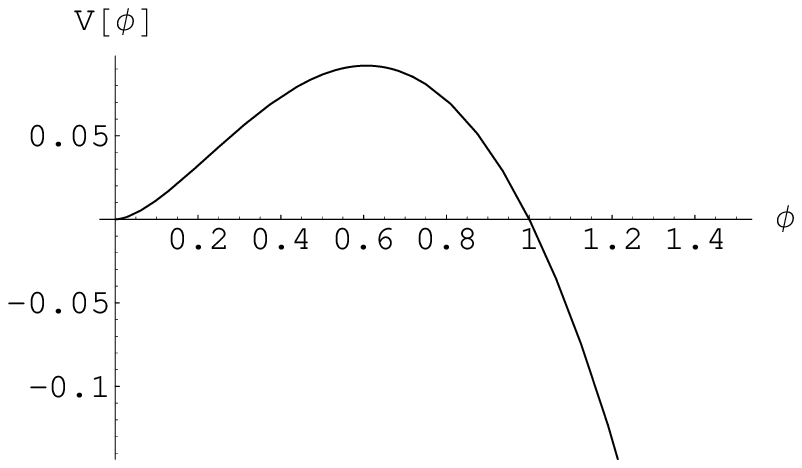}
\includegraphics[width=4.8cm]{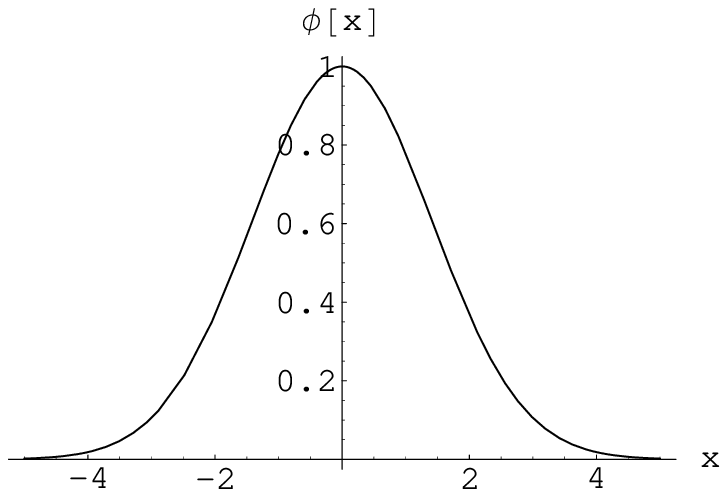}}
\caption{The shape of the potential $V(\phi)$ and of the lump solution.}
\end{figure}
\begin{equation}
\label{unosue}  
V(\phi) = - {{1}\over{4}} \phi^2 {\mathop{\rm ln}} \phi^2.
\end{equation}
It has a maximum at $e^{- 1/2}$, corresponding to the D25-brane,
and a local minimum at $0$, 
corresponding to the closed string vacuum, 
where the curvature diverges. The last fact beautifully mimics 
the idea of decoupling of the open string degrees of freedom 
in the stable vacuum, which is a feature of the standard tachyon condensation.
The potential admits a Gaussian lump solution independent of the $y$ variables 
\begin{equation}
\label{lump}
 \bar{\phi} (x) = \exp\Big[- {{x^2}\over{4}}\Big], 
\end{equation}
shown in Fig.1, that represents a D24-brane.
After $\phi \rightarrow \bar{\phi} + \phi$, making the ansatz 
\begin{equation}
\label{expa}
\phi = \sum_n \bar{\phi}_n (y) \psi_n (x),
\end{equation} 
the computation of the  
spectrum of fluctuations around the lump
is effectively reduced to the one dimensional Schr\"odinger equation for the
harmonic oscillator

\begin{equation}
\label{schroe}
- {{d^2 \psi}\over{dx^2}} + 
\bigg[ -\frac32 +\frac14 x^2 \bigg] \psi (x) = m^2 \psi (x),
\end{equation}
that determines  in this way the ``open string'' 
spectrum $m^2 = n-1$ for $n\geq0$. 

The advantage of the  model is
that we know the minimum to be at the value where the original tachyon field
$\phi$ is equal to $0$, therefore the exact values of the condensates for the 
whole tower are obtained by solving the following equation:

\begin{equation}
\label{determ1}
0 = \bar\phi (x) + \sum_n \bar\phi_n \psi_n (x) .
\end{equation} 

Since the Gaussian itself is 
the first eigenfunction of the harmonic oscillator, the solution of 
the equation is immediately found. It simply amounts 
to fixing the expectation value $-1$ for the tachyon ($n=0$), 
and zero for all the other excitations. 
This means that only the tachyon condenses.

We want to use this model  
to describe the decay of a D-brane due to the overcritical electric field. 
The decay of the D-brane is now driven by the infinite 
tower of former massive fields now being tachyonic. We will see that it is still
possible to study the decay of a D24-brane using the potential for the tachyon
of a D25-brane.

Modifying the model in the presence of an electric field,  
simply amounts to use for the $y$ coordinates the open string metric, and 
simultaneously to introduce for them the star-product\footnote{We will again 
always consider here either effective one dimensional $x$-variable 
problems, or static or linearized ones in the $y$ variables, 
therefore the star product will never play any role, 
and we will omit to write it.}.

Then, instead of (\ref{cubic}), the action becomes 

\begin{equation}
\label{overcubic}
S = {{1}\over{g_o^2}} \sqrt{1-E^2} \int d^{25}y \, 
dx \bigg[-{{1}\over{2}} G^{\mu \nu} \partial_{\mu} 
\phi \partial_{\nu} 
\phi - {{1}\over{2}} {(\partial_x \phi )}^2 - V(\phi)\bigg],
\end{equation}
where we included the factor $\sqrt{1-E^2}$ in order to
recover the correct D-brane tension. Technically it would be equivalent to
take into account the factor $\sqrt{-\mbox{det}G}=\vert 1-E^2\vert$ in the
measure and to replace $g_o^2$ by
$g_o^2\sqrt{-\mbox{det}(g+B)}=g_o^2\sqrt{1-E^2}$, as is done for the DBI
action in ref.\cite{sw}. Note that we insist on keeping the string coupling
real for use in full String Field Theory. The action (\ref{overcubic})
is thought to be an effective one for the lowest string excitation.  
Having this in mind, the presence of an imaginary
factor in (\ref{overcubic}) is no obstacle and in agreement with the DBI
analysis.    

The ``transverse'' part in the $x$ variable is not touched by the 
electric field. In particular, the same lump (\ref{lump}) is still a 
solution of the new equations of motion. 
With $G_{\mu\nu}$ from Sec.2, the action (\ref{overcubic}) can be written as
\begin{eqnarray}
\label{explovercubic}
&&S = {{1}\over{g_o^2}}  {{1}\over{\sqrt{1-E^2}}}
\int d^{25}y \,
dx \bigg[{{1}\over{2}} {(\partial_{0} 
\phi)}^2 - {{1}\over{2}} {(\partial_1 \phi )}^2 
\nonumber \\
&&- {{1}\over{2}} (1-E^2)\sum_a {(\partial_a \phi )}^2 - {{1}\over{2}} 
(1-E^2) {(\partial_x 
\phi )}^2 - (1-E^2)V(\phi) \bigg].
\end{eqnarray}
Here, $a$ labels directions parallel to the brane, but perpendicular to the 
electric field.

Referring ourselves to the action (\ref{explovercubic}),  
we first collected an overall factor 
in front, in order to make the following analysis clearer. We
are interested in determining the solutions of the classical
equations of motion, and to discuss their stability. The overall factor 
does not change this kind of analysis, since it does not alter the 
equations 
of motion and it does not affect the issue of stability around the extrema
of the classical potential.
We can therefore 
disregard this factor as far as this analysis is concerned. We will
return later on its actual role. 

From inspection of (\ref{explovercubic}), 
one can see that  
the time derivative is normalized as in the $E = 0$ case, 
but the potential has 
a factor in front with respect to that case, which changes sign in the 
overcritical case. We realize therefore that, together with the already 
known effect of the extra contribution coming from transverse 
motion\footnote{Compare the 
derivatives in the direction $y^a$ in (\ref{explovercubic}) with
the contribution from 
$k_\perp$ in (\ref{spectrumel}) and (\ref{spectrumel2}).}, the net 
effect of the overcritical electric field 
is to reverse the sign of the potential.
This potential is drawn in Fig.2. 
\begin{figure}
\centerline{\includegraphics[width=6cm]{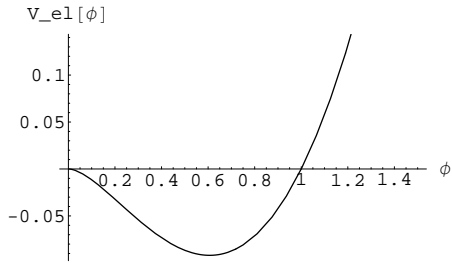}}
\caption{The shape of the potential $V_{el}$ up to a factor $|1-E^2|$.}
\end{figure}
Such an effect is consistent with the general argument concerning the
sign of the squared masses  
in the overcritical electric background. 
The reversed tachyon potential 
for static configurations has a stable minimum at $e^{-1/2}$, 
that corresponds to the D25-brane. This means that
the tachyon
is now a stable (positive mass) state in the spectrum of the 
D25-brane\footnote{The D25-brane is affected as well as the D24-brane
(lump) by the electric field, which is switched on in directions
common to both of them.}.     

The model is therefore suitable to mimic the situation in the full 
String Field Theory, and one can be further convinced of this by 
examining the expanded action around the lump solution by replacing $\phi$
with $\bar{\phi} + \phi$. 
The action reads now
\begin{eqnarray}
\label{overcubicexp}
S&=& 
 {{1}\over{g_o^2}} \sqrt{1-E^2} \int d^{25}y \, dx ~\Big[ - {{1}\over{2}} 
{\Big({{d\bar\phi}\over{dx}}\Big)}^2 - V(\bar\phi)\Big] \nonumber\\
&& + {{1}\over{g_o^2}} 
\frac{1}{\sqrt{1-E^2}} \int d^{25}y \, dx ~\Big[{{1}\over{2}} {(\partial_{0} 
\phi)}^2 - {{1}\over{2}} {(\partial_1 \phi )}^2 
- {{1}\over{2}} (1-E^2)\sum_a {(\partial_a \phi )}^2 \nonumber \\
&& - {{1}\over{2}} (1-E^2) \phi \big( - \partial^2_x + V''(\bar\phi)\big)\phi
+ \ldots
\Big].
\end{eqnarray}
The lump solution is still interpreted as a codimension $1$ brane. 
Its asymptotes 
at $x=\pm \infty$ correspond to the absence of the D25-brane. 
The first two terms correctly reproduce the D24-brane 
tension with the 
Born-Infeld factor which accounts for the electric background. This factor 
becoming imaginary was the object of the discussion in Sec.2. 

The remaining part gives just the effective action for the 
fluctuations, which 
is used to determine the dynamics of the decay through the 
equations of motion 
and the analysis of stability, which are not influenced by the overall 
factor.  
The effective one dimensional Schr\"odinger problem is untouched by the 
presence of the electric field, the spectrum of fluctuations is therefore the 
same. But when we plug back the mode expansion in the action, we see that the 
eigenvalues of the Schr\"odinger equation contribute with a 
multiplicative factor $(1-E^2)$. This is precisely the modification of the 
squared mass for the open string spectrum according to the general treatment
in Sec.2. We see here that the theory living on the 
worldvolume of the lump 
suitably describes the infinite tachyonic tower of states of
an electrified D-brane. 
After integration over $x$, the resulting
multiscalar potential can be taken as a toy model for the behaviour of the
full String Field Theory in the presence of an overcritical electric field.

We can now predict inside this toy model the value of the condensate for the 
infinite tachyonic tower. The final stage representing the end 
of the rolling manifests itself in this toy model as a stable minimum 
of the reversed potential (corresponding to the old maximum)
\footnote{Of course it is open whether this feature is shared by the real  
String Field Theory potential. At this stage we take it as a conjecture.}.

We can apply the same technique as in the 
case without electric field to compute the 
expectation values, only requiring that the lump profile plus 
the fluctuations reduces the original tachyon to its true minimum at 
$e^{- 1/2}$.

The overcritical counterpart of the equation (\ref{determ1})
is therefore
\begin{eqnarray}
\label{master}
e^{- 1/2} & = &  \bar{\phi}(x) + \sum_{n=0}^{\infty} \bar\phi_n \psi_n(x)\\
& = & \exp[-{{x^2}/{4}}] + \sum_{n=0}^{\infty} \bar\phi_n \psi_n(x), \nonumber 
\end{eqnarray}
where the eigenfunctions $\psi_n (x)$ are the usual harmonic oscillator basis
constructed in terms of Hermite polynomials
\begin{equation}
\label{Herm}
\psi_n (x) = {{1}\over{2^{n/2} \sqrt{n!}}}\, H_n \Big({{x}\over{\sqrt{2}}}\Big) \,
\exp[-{{x^2}/{4}}].  
\end{equation}
If we absorb the Gaussian profile in a redefinition of the coefficient 
$\bar\phi_0$, then Eq.(\ref{master}) represents an expansion of a constant in 
terms of a Hilbert space  basis. A constant is certainly not in the 
Hilbert space. Still it is possible to compute the coefficients
$\bar\phi_n$ by performing the scalar product with any elements of the 
basis\footnote{The basis is normalized such that
$\int dx \, \psi_n (x) \, \psi_m (x) 
= \sqrt{2\pi} \delta_{n,m}$.}. 
From 
\begin{equation}
\label{scalpro}
\bar\phi_n = {{1}\over{\sqrt{2\pi}}} \int_{-\infty}^{\infty} dx \, \psi_n (x) 
\Big(e^{- 1/2} -
\exp \Big[-{{x^2}\over{4}}\Big]\Big),
\end{equation}
one gets 
\begin{equation}
\label{condens}
 \bar\phi_{2n+1} = 0,~~~~~~ \bar\phi_{2n} = - \delta_{n,0} + {{2^{- n}}\over{n!}} \sqrt{{{2 \,(2 n)!}\over{e}}}.
\end{equation}
This is the value of the condensates for the whole tower 
of states at the true vacuum.
At large level number $n$, using Stirling's approximation, one gets a behaviour
\begin{equation}
\label{behav}
\bar\phi_{2n} \longrightarrow \sqrt{{{2}\over{e}}} {(\pi n)}^{- {{1}\over{4}}}.
\end{equation}

Due to the slow decrease of $\bar{\phi}_{2n}$ in (\ref{behav}),
the resulting series is not absolutely and not uniformly (for all $x$)
convergent. However, the oscillations with $n$ of the sign of the Hermite
polynomials enforce uniform convergence in finite $x$-intervals. It is
straightforward to check this at least numerically.

Although the expectation values for higher modes go to zero with 
$n \rightarrow \infty$ the decrease is not as fast as in the case without 
electric field \cite{z, mz}. In accordance with our previous discussion we
take the {\it slow} decrease as an indication that level truncation would be a 
bad approximation. 
Indeed, we performed some explicit checks in level truncation,
and they confirmed these expectations. Taking into account up to level
$n=2$, and expanding the reversed effective potential up to
the fifth power in these scalars, a minimum is found at ${\bar\phi}_0 = 
-0.226$, ${\bar\phi}_1 = 0$, $\bar\phi_2 = 0.555$. This has to be 
compared with the exact result (\ref{condens}), namely ${\bar\phi}_0 = 
-0.142$, ${\bar\phi}_1 = 0$, $\bar\phi_2 = 0.607$. At first sight 
this may seem not so bad an approximation, but we notice that, first, 
truncating at even powers up to the fourth, sixth and eighth ones  
produce instead no minimum, and, 
second, 
truncation to seventh and ninth powers produce a worse minimum. For example, 
with the seventh powers included the minimum is at ${\bar\phi}_0 = 
-0.0831$, ${\bar\phi}_1 = 10^{-8}$, $\bar\phi_2 = 0.350$, 
and with the ninth powers included the minimum is at ${\bar\phi}_0 = 
-0.0463$, ${\bar\phi}_1 = 10^{-4}$, $\bar\phi_2 = 0.259$. If 
convergence of level truncation is eventually to be obtained, it is done with
wide oscillating behaviour, similar to the convergence of the sum in (\ref{master}). 
  
A last comment concerns the nature of the endpoint of the condensation
process. The toy model setup is designed to simulate the decay of a 
D24-brane as a lump solution of the D25 tachyon field $\phi$. For the case
of undercritical electric field the minimum of the potential for $\phi$
corresponds to the absence of the D25. Now in the overcritical case
$\phi$ takes a value as in the presence of a D25.  
Within the lump-based model one has no possibility to decide 
whether this value for $\phi$ indicates a D25 or the true vacuum in which
$\phi$ could assume just the same value ($\phi$ does not drive a D25 decay). 
However, beyond the model 
one should expect that it will be the duty of the tower of
unstable D25 modes to prevent a formation of a D25.

From the open string point of view, the analogy with pair production of
pointlike charges suggests that the endpoint of the condensation process
corresponds to the discharge of the overcritical electric field: dipoles
nucleating from the vacuum easily stretch to infinity until they screen
the electric field making it critical or undercritical \cite{sbranes2}.
When looking at the
D-brane decay, this means that the fluctuations around the unstable
configuration are going to lower the value of the electric background
itself towards critical or undercritical values. In the T-dual picture,
one expects therefore that the T-dual superluminal brane (S-brane) would lower
its velocity to (under-) luminal, thereby becoming a usual timelike 
D-brane. This would
correspond to what is observed for example in \cite{sbranes1}, where the
final stage of the S-brane evolution is represented by a flattening of its
profile. This gives a hint that a remnant could be left after condensation,
in accord to the general expectation of fundamental string fluxes as final
products of the decay in the presence of electric fields \cite{mukho}. In
order to get a full description of this phenomenon one would have to allow
dynamics for the electric field as well and properly consider
backreaction of the D-brane. In this respect we think that a 
closer connection to the
S-brane picture could again be fruitful.

\section{Acknowledgments}
We thank Antonio Bassetto, Loriano Bonora, Giancarlo De Pol, 
Matthias Gaberdiel, Carlo Maccaferri, Adriano Parodi, Rodolfo Russo, 
Roberto Valandro and Hyun Seok Yang for useful discussions. 
We would like to thank Koji Hashimoto for pointing out to us
the connection with S-branes, and for very interesting e-mail exchange.
DFG supported 
H.D. and A.T. within the ``Schwerpunktprogramm Stringtheorie 1096'' 
and  M.S. under the project SA 1356/1.

\end{document}